\documentclass{kluwer}    % Specifies the document style.
\usepackage{psfig}
\newdisplay{guess}{Conjecture}
\begin{document}

\begin{article}
\begin{opening}

\title{Gamma-Hadron Separation using \v Cerenkov Photon Density Fluctuations}

\author{V. R. \surname{Chitnis} and P. N. \surname{Bhat}\thanks{pnbhat@tifr.res.in}}
\runningauthor{Chitnis and Bhat}
\runningtitle{Gamma-Hadron Separation using \v Cerenkov Photon Density Fluctuations}
\institute{Tata Institute of Fundamental Research,\\
 Homi Bhabha Road,
 Mumbai 400 005, India.}
\date{}

\begin{abstract}
   In the atmospheric \v Cerenkov technique $\gamma-$ rays are detected
against the abundant background produced by hadronic showers. In order to
improve the signal to noise ratio of the experiment, it is necessary to reject
a significant fraction of hadronic showers. Traditional background rejection 
methods based on image shape parameters have been extensively used for the 
data from imaging telescopes. However, non-imaging \v Cerenkov telescopes have 
to develop very different means of statistically identifying and removing 
cosmic ray events. Some of the parameters which could be potentially important
for non-imaging arrays are the temporal and spectral differences, the lateral
distributions and density fluctuations of \v Cerenkov photons generated by
$\gamma-$ ray and hadron primaries. Here we study the differences in 
fluctuations of \v Cerenkov photon density in the light pool at the observation 
level from showers initiated by photons and those initiated by protons or
heavier nuclei. The database of simulated events for the PACT array has been
used to evaluate the efficiency of the new technique. Various types of density 
fluctuations like the short range and medium range fluctuations as well as
flatness parameter are studied. The estimated quality factors reflect the 
efficiencies with which the hadrons can be rejected from the data.  Since some
of these parameters
are independent, the cuts may be applied in tandem and we demonstrate
that the proton rejection efficiency of $\sim$ 90\% can be achieved.  
Use of density fluctuations is particularly suited for wavefront sampling
observations and it seems to be a good technique to improve the signal to
noise ratio.

\end{abstract}

\end{opening}

\section{Introduction}
The atmospheric \v Cerenkov technique has become an increasingly mature 
experimental method of very high energy (VHE) $\gamma$-ray astronomy in the 
recent years \cite{we88,cr93,cw95,aa97,on98,cw99}. Major effort has gone into
the optimization of $\gamma$-hadron separation\cite{fe97,vc01}, the energy
calibration, energy resolution \cite{ho00} and the evaluation of spectra
of $\gamma$-ray sources. Both imaging and non-imaging techniques play an
increasingly important role in measuring the precise spectra of $TeV~\gamma$-ray
sources over a wide energy range \cite{ah00,ab01,ob00,nh01}. Among the 
non-imaging telescopes are the detector arrays based on solar concentrators
like the CELESTE, GRAAL and STACEE have the potential to achieve unprecedented
low energy thresholds for ground based $\gamma$-ray detectors \cite{cel96,arq97,on95}. The imaging technique is also evolving with several large (10 $m$ class)
imaging telescopes forming a  stereoscopic array to achieve unprecedented
angular resolution $e.g.$ HESS\cite{ah97}, VERITAS{\cite{we96}}. These are 
currently under development. A single large imaging telescope (17 $m$ 
diameter) is also under development which could achieve the lowest ever
threshold of $\sim 10~GeV$ for primary $\gamma$-rays \cite{bl98}.

However, there has been a major difficulty in applying the atmospheric
\v Cerenkov technique successfully for $\gamma -$ ray astronomy.  The abundant
charged cosmic ray particles generate \v Cerenkov light akin to that produced
by the $\gamma -$ rays as a result of which the $\gamma -$ ray signal is buried
in a vast sea of cosmic ray background. Accurate location of the direction from
which the primary particle is incident at the top of the atmosphere has been
an important feature of successful observations since hadronic showers are
isotropic. This has been the motivation to develop very good angular resolution
capabilities \cite{mab02}. But since $\gamma $-ray sources are weak one needs
additional methods to reject the on-axis hadronic showers from the data.
Such techniques have been developed for imaging telescopes \cite{fe97} while
they are  still being developed for non-imaging telescope arrays \cite{vc01}.

The best distinction between $\gamma $-ray and proton showers should be based 
on an ideal parameter that does not show large deviations from its mean
value $i.e.$ it has a narrow distribution, and also whose fitted $\gamma $-ray 
curve should be well separated from the corresponding one for the hadronic
showers. Often this is hardly achievable by using a single parameter and one 
often uses two or more parameters in tandem \cite{vc01} or simultaneously  uses 
several parameters together with their correlations \cite{ack89} to achieve
the same goal. Major difficulty is that these parameters often vary with
primary energy and core distance, hence the separability too.

Methods for the efficient discrimination of photon and hadron initiated 
showers have been derived from the differences in the intrinsic properties
of the \v Cerenkov radiation from pure electromagnetic and hadronic 
cascades.  Differences manifest in the spatial as well as temporal 
distributions of the \v Cerenkov photons at the observation level. As a result,
systematic studies of these photons as received at the observation level could
lead to the development of techniques to distinguish between hadronic or photon
primaries. There are several Atmospheric \v Cerenkov arrays designed precisely
to apply these techniques to ground based VHE $\gamma -$ ray astronomy
\cite{cel96,arq97,pnb98,on96,oc97,tum90}.

The technique of separating proton or heavy nuclei initiated showers from
$\gamma $-ray initiated showers was successfully developed first using
atmospheric \v Cerenkov images \cite{fe97,tcw96}. 
A new method is based on the image surface brightness which seems to change with
primary energy and species. This shows the importance of the photometric 
information of \v Cerenkov images in addition to the image shape parameters    
\cite{baw97} demonstrating that the photon densities too contain
species dependent signature. More recently, a method based on the differences
in the fluctuations of light intensity in the images of showers initiated by
$\gamma $-rays and cosmic ray hadrons has been developed to offer additional
background rejection capability \cite{bpv02}.

In the present work we similarly explore a new discrimination technique from a
study of the spatial profile of \v Cerenkov light brightness from 
pure electromagnetic cascades as well as hadronic cascades generated by
very high energy primaries. The technique is based on the differences in the 
intrinsic fluctuations in the \v Cerenkov photon densities at the observation 
level produced in the pure electromagnetic and hadronic cascades. The photon 
density fluctuations are classified in terms of their spatial extent, {\it 
i.e.} `short', `medium' and `long' range and then parameterized. We then 
investigate the sensitivity of these three basic parameters to primary species.
The relative merits of these parameters are compared in terms of the quality
factors which are indicators of the efficiency with which showers of hadronic
origin could be rejected in order to improve the signal to noise ratio of the
data.

In \S 2 of this paper, details of simulations are given, followed by the 
definition of the figure of merit for discrimination between $\gamma-$ rays and
cosmic rays in \S 3. In \S 4,  we define the three different parameters based
on relative photon density fluctuations and study their core distance
dependences. Then we present the quality factors derived for each of the
parameters for various primary energies. The dependence of the quality factors
on the telescope opening angle and incident angle of the primary at the top
of the atmosphere are discussed in \S's 5  \& 6 respectively. The \S7
contains results on the species dependence of the quality factors. A brief
discussion of the results is presented in \S 8 and conclusions are summarized
in \S 9.

\section{Simulations}

CORSIKA (version 5.60), \cite{kn98,hec98} has been used to
simulate \v Cerenkov light emission in the earth's atmosphere by the secondaries
of the extensive air showers generated by cosmic ray primaries or $\gamma-$ 
rays. This program simulates interactions of nuclei, hadrons, muons, electrons
and photons as well as decays of unstable secondaries in the atmosphere. It
uses EGS4 code \cite{ne85} for the electromagnetic component of the air shower
simulation and the dual parton model for the simulation of hadronic interactions
at $TeV$ energies. The \v Cerenkov radiation produced within the specified band
width (300-650 $nm$) by the charged secondaries is propagated to the ground. 
The US standard atmosphere parameterized by Linsley \cite{us62} has been used.
The position, angle, time (with respect to the first interaction) and
production height of each photon hitting the detector on the observation level
are recorded.

\begin{figure}
\centerline{\psfig{file=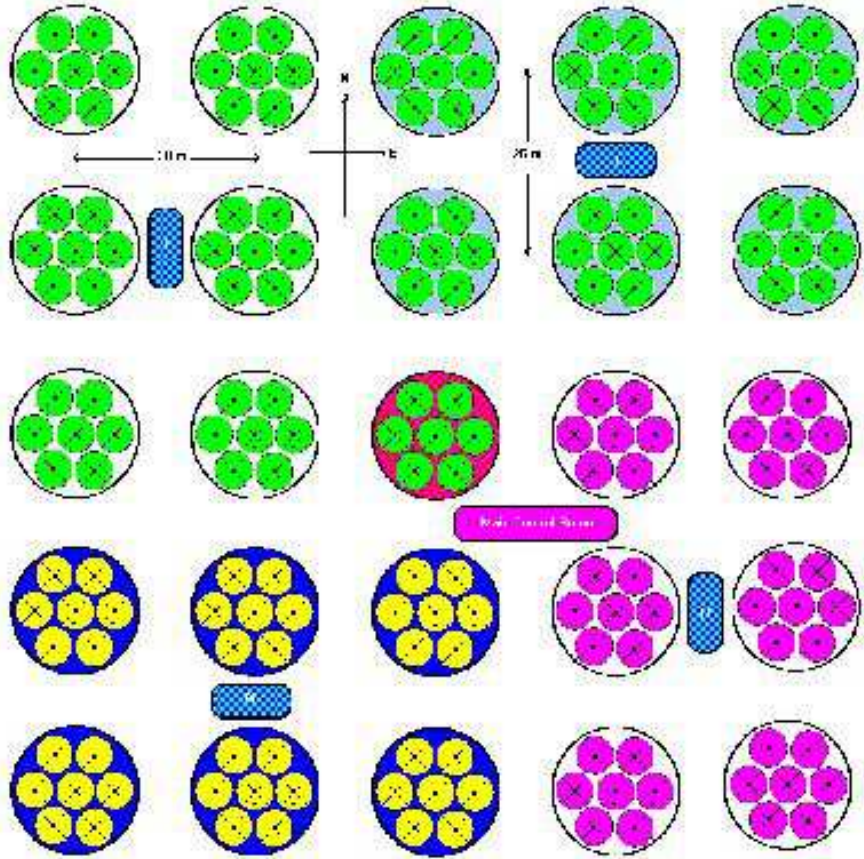,height=12cm,width=12cm}}
\caption{The PACT array showing the approximate positions of the 25 telescopes
covering an area of about $80m\times~100m$.
} \label{array_fig}
\end{figure}

       In the present studies we have used Pachmarhi (longitude:
78$^{\circ}$ 26$^{\prime}$ E, latitude: 22$^{\circ}$ 28$^{\prime} N$ and
altitude: 1075 $m$) as the observation level where an array of \v Cerenkov
detectors each of area\footnote{This is the total reflective area 
of 7 parabolic mirrors of diameter 0.9 $m$ deployed par-axially on a single 
equatorial mount.} 4.35 $m^2$ is deployed in the form of a rectangular
array. We have assumed 17 detectors in the E-W direction with a separation of
25 $m$ and 21 detectors in the N-S direction with a separation of 20 m, making
a total of 357 detectors deployed over an area of $400m\times~400m$.  This
configuration, similar to the Pachmarhi Array of \v Cerenkov Telescopes (PACT;
figure \ref{array_fig})\cite{pnb98} but much larger, is chosen so that one 
can study the core distance
dependence of various observable parameters. Primaries consisting
of $\gamma-$ rays, protons, $He$ and iron nuclei incident vertically on the top
of the atmosphere 
are simulated in this study and have a fixed core position which is chosen
to be the detector at the centre of the array. The resulting \v Cerenkov pool
is sampled by all the 357 detectors which are used to study the core distance
dependence of the parameters studied here. All the telescopes are assumed to
have their optic axes aligned vertically. From practical considerations PACT
has been divided into 4 sectors each of 6 telescopes. The physical size of a
sector is approximately $20~m\times 50~m$ or $25~m\times 40~m$ depending on the
orientation. In simulation however, there are 56 such sectors or 12 PACT like
arrays (comprising 4 sectors) in the larger simulated array of
357 detectors. These are used for computing the density based parameters and
study their core distance dependences.

An option of variable bunch size of the \v Cerenkov photons is available
in the package which serves to reduce the requirement of hardware resources.
This basically defines a maximal number of \v Cerenkov photons that are
treated together as single entity.
However since we are interested in the fluctuations of each of the estimated
observables, we have tracked single photons for each primary at all energies.
Multiple scattering length for electrons and positrons is decided by the
parameter STEPFC in the EGS code which has been set to 0.1 in the present
studies \cite{fo78}.  The wavelength dependent absorption of \v Cerenkov photons
in the atmosphere is not however taken into account.  The present conclusions
are not expected to be dependent on photon wavelengths\cite{rba01}. 

\section{Figure of merit of a parameter}

The figure of merit of a parameter that can distinguish between VHE
$\gamma -$rays
and cosmic ray hadrons depends primarily on two factors. Firstly, it should
accept most of the $\gamma -$rays and secondly it should be able to reject most
of the hadrons.  We define such a figure of merit which is often called as
{\it quality factor}, as \cite{vc01,rb98}:

\begin{equation}
q={{N_a^{\gamma}} \over {N_T^{\gamma}}} \left( {{N_a^{cr}} \over {N_T^{cr}}} \right) ^{-{1 \over 2}}
\end{equation}
  
where $N_a^{\gamma}$ is the number of $\gamma$ rays accepted, 
$N_T^{\gamma}$ is the total number of $\gamma$ rays,
$N_a^{cr}$ is the number of background cosmic rays accepted and 
$N_T^{cr}$ is the total number of background cosmic rays.

The quality factor thus defined is independent of the actual number of 
$\gamma -$rays and protons recorded. This is slightly different from the 
Q-factor generally used to measure improvement in the signal to noise ratio
after applying cuts based on \v Cerenkov image shapes and orientation 
\cite{baw97}.

\section{Types of fluctuations}
\subsection{Local Density Fluctuations}
Local density fluctuations (LDF) are defined as the ratio of the RMS variations 
to the mean number of photons in the 7 mirrors of a telescope. In the particular
configuration chosen in the present study, the 7 mirrors form a compact pattern.
Hence LDF, in this context, represents the short range ($\sim~1~m$) photon 
jitter. Here we
try to compare LDF's for $\gamma $-ray primaries {\it vis-a-vis} hadronic
primaries and see if there is any tangible difference which could in-turn be
used to discriminate the hadronic background.

\subsubsection{Lateral Distribution of LDF}

\begin{figure}
\centerline{\psfig{file=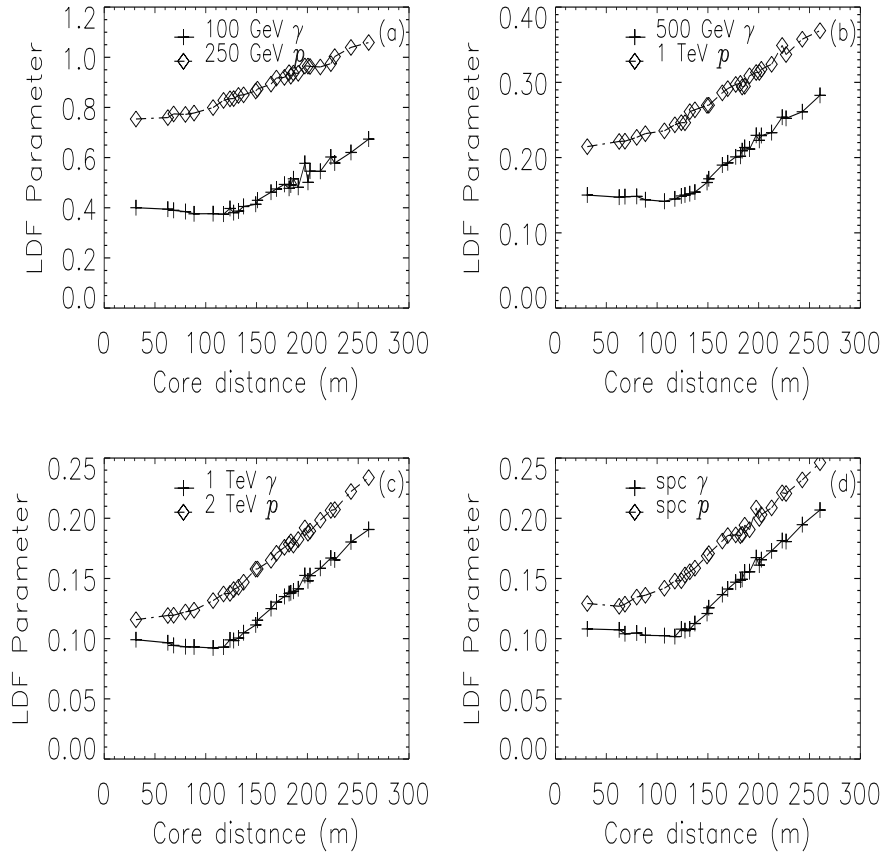,height=12cm,width=12cm}}
\caption{
Radial variations of LDF averaged over 100 showers each for $\gamma $-ray 
(plus) and proton 
(diamonds) primaries of energies (a) 100 $GeV$ \& 250 $GeV$ (averaged over
300 showers each) (b)  500 $GeV$ \& 1
$TeV$ and (c) 1 $TeV$ \& 2 $TeV$ respectively. Each set of energies for
$\gamma $-ray and proton energies are chosen such that they have comparable
\v Cerenkov yield. Panel (d) shows similar distributions for $\gamma $-rays
\& protons of primary energies randomly chosen from a power law spectrum of 
slope -2.65 (see text for details).} \label{rad_ldf}
\end{figure}

We have computed the LDF for each of 357 detectors for $\gamma $-ray and proton
primaries whose energies at the top of the atmosphere are chosen such that 
they have comparable \v Cerenkov yield on the ground. Figure \ref{rad_ldf}
shows the radial
variations of mean LDF both for $\gamma $-rays and protons for three pairs of
primary energies as shown. The panels $a,~b$ and $c$ show the radial variation 
of LDF for mono-energetic primaries while panel $d$ shows the same when the
energies of the primary are picked randomly from a differential power law 
spectrum (within the energy band of 0.5 - 10 $TeV$ for $\gamma $-rays and
1 - 20 $TeV$ for protons) of slope -2.65. One can readily see that LDF 
increases smoothly with increasing core distance for both $\gamma $-ray 
and proton primaries beyond the hump distance of about 130 $m$ at
this observation level \cite{vc98}. LDF for proton primaries is distinctly
higher than that for $\gamma $-rays at all primary energies considered here
while their absolute values as well as their separations decrease with
increasing primary energy as expected \cite{vc99}.

\subsubsection{Quality Factors using LDF}

\begin{figure}[ht]
\centerline{\psfig{file=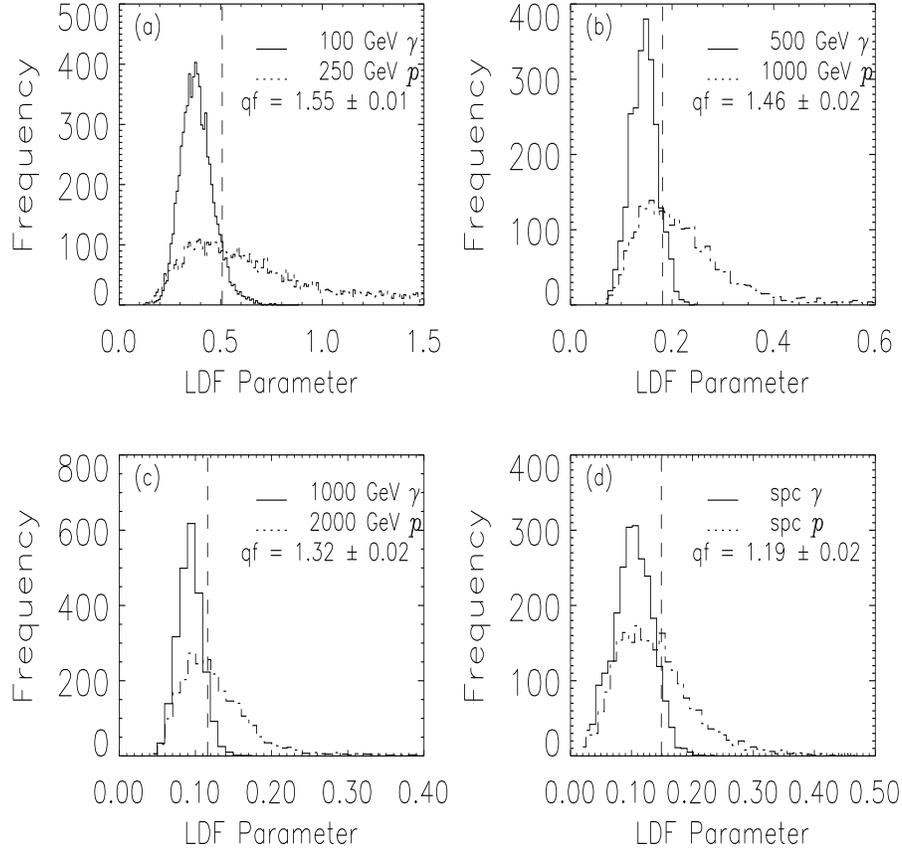,height=12cm,width=12cm }}
\caption{
Distributions of LDF for $\gamma-$ray and proton (dotted line) primaries of 
three sets of energies as in figure \ref{rad_ldf}. The dashed vertical lines
show the threshold values chosen to yield maximum quality factors in each case.
The number of showers used are same as those mentioned in figure \ref{rad_ldf}.}
\label{qf_ldf}
\end{figure}

Figure \ref{qf_ldf} shows the distributions of LDF's for $\gamma-$ray and 
proton primaries of various energies incident vertically at the top of the
atmosphere.  The distributions for proton primaries are more skewed compared to
those of $\gamma-$rays and hence result in good quality factors. The widths of
the distributions for both types of primaries decrease with increasing energy
and their relative separations too decrease. The vertical dashed lines show the
threshold values chosen to yield maximum quality factors and retain atleast
30\% $\gamma-$ray signal. Only detectors within a core distance of 150~$m$ are
used in these distributions.
 
Table \ref{tab_ldf} shows the quality factors estimated using LDF as a parameter
for the same primary energies of $\gamma-$rays and protons shown in figure 
\ref{rad_ldf}.  The second column shows the threshold LDF values such that the
showers whose LDF values are above this are discarded yielding a quality
factor shown in column 3. The column 4 lists the fraction of the $\gamma-$ray
and proton events retained in the process. The
quality factors decrease  steadily with increasing primary energies probably
because the intrinsic fluctuations are smoothed out at higher primary energies.
This is consistent with the radial variations shown in figure \ref{rad_ldf}. 
In addition,
the threshold LDF values decrease with increasing primary energies since the
value of LDF's also decrease with energy.

\begin{table}
\caption{Quality factors derived using local density fluctuations for primaries
of various energies incident vertically at the top of the atmosphere.}
\label{tab_ldf}
\vskip 0.25cm
\begin{tabular}{llllc}
\hline
Primary & Threshold  &  Quality Factor  &  Fraction  \\
Energy  & LDF        &                  &  Accepted  \\
        &            &                  &  \\
\hline
\hline
100 $GeV$ $\gamma-$rays &  0.51  & 1.55 $\pm$ 0.01  & 0.922  \\
and 250 $GeV$ protons    &&&0.355\\
\hline
500 $GeV~\gamma-$rays & 0.18  & 1.46 $\pm$ 0.02 & 0.880  \\
and  1 $TeV$ protons &&&0.364\\
\hline
1 $TeV~\gamma-$rays &  0.12 & 1.32 $\pm$ 0.02 & 0.872 \\
and 2 $TeV$ protons &&&0.435\\
\hline
Spectrum of $\gamma-$rays & 0.15 & 1.19 $\pm$ 0.02 & 0.91 \\
and protons             &      &                 & 0.59 \\
\hline
\end{tabular}
\end{table}

The lateral distributions of $\gamma-$rays and hadrons are
distinctly different within the hump distance of about 130 $m$ \cite{vc98}. The 
$\gamma $-ray lateral distribution is relatively flat until the hump distance
and then falls while that for proton primaries it falls monotonically right
from the core. This difference in the lateral profile could very well give rise
to differences in the LDF values for the two primaries as shown in figure 
\ref{qf_ldf}. Hence LDF could arise purely out of the Poisson fluctuations of
the number of \v Cerenkov photons incident on each of the 7 mirrors. When the 
quanity ${1}\over{\sqrt{n(r)}}$ is plotted as a function of core distance, 
(where $n(r)$ is the number of \v Cerenkov photons incident on a mirror at a
core distance $r$) it coincided with that shown in figure \ref{rad_ldf}
demonstrating that LDF does not contain any non-Poissonian fluctuations. Hence
the modest quality factors arise due to the differences in the lateral
distributions.  This shows that the kinematical differences in the cascade
development by primary $\gamma-$rays and protons give rise to only long range
fluctuations.

\subsection{Medium Range Density Fluctuations}

Medium range density fluctuation (MDF) is defined as the ratio of the RMS 
variations of the total number of photons detected in each of the 6 
telescopes\footnote{which is the sum of the photons incident on all the 7 
mirrors constituting the telescope.}in a sector to the average number of photons
incident on a telescope. In other words, MDF is a measure of the variation of
\v Cerenkov photons over a medium range of $\sim 50~m$. As mentioned before,
we have 56 independent sectors in the large array chosen in our simulation,
situated at various core distances. 

\subsubsection{Lateral Distribution of MDF}

\begin{figure}[ht]
\centerline{\psfig{file=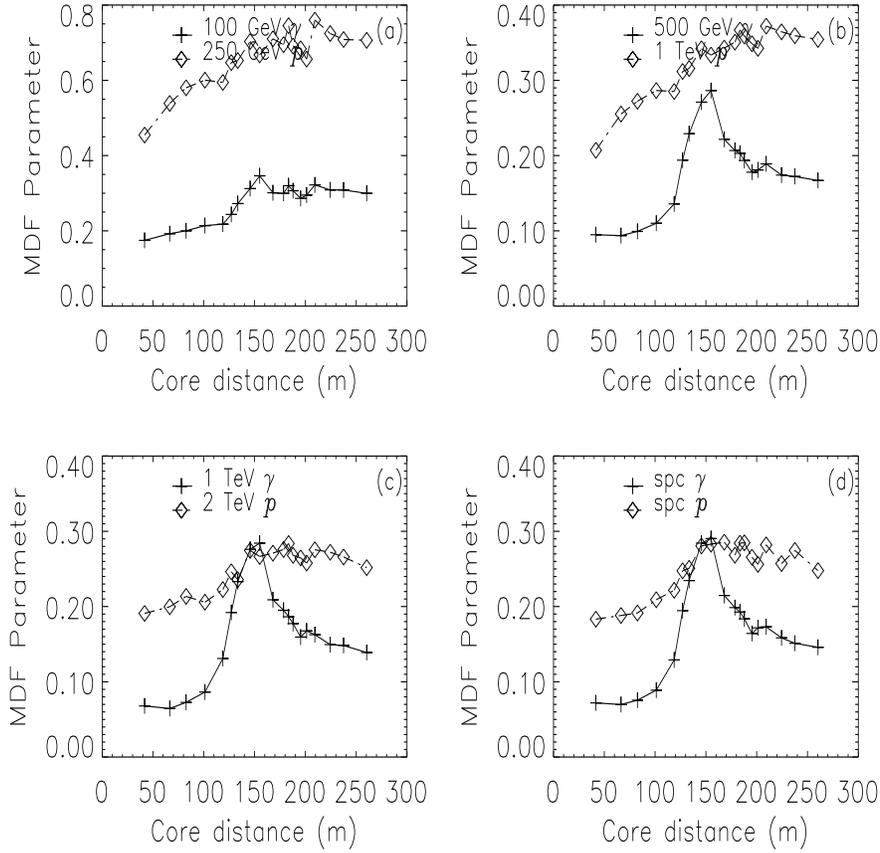,height=12cm,width=12cm}}
\caption{
Radial variations of MDF for $\gamma $-ray (plus) and proton (diamonds)
primaries. The rest of the details are same as in figure \ref{rad_ldf}. The
number of showers used are 300 each in panel $a$ and $b$ while it is 100 each 
for $\gamma-$rays and protons for the rest.} \label{rad_mdf}
\end{figure}

Figure \ref{rad_mdf} shows the radial variation of MDF for the same four pairs 
of energies for $\gamma-$rays and protons as in the case of LDF. It can be
seen that the MDF values of $\gamma-$rays and protons are separated at lower
primary  energies showing that it can well be used as a parameter to distinguish
between them. The MDF values for proton primaries are less sensitive to the core
distance as compared to LDF values at almost all primary energies. However the
$\gamma $-ray primaries exhibit a prominent peak around the hump region
especially at higher energies. The maximum MDF could exceed that for protons at
energies $\ge ~2~TeV$. The absolute values of MDF decrease with increasing
primary energies as in the case of LDF.
 
\subsubsection{Quality Factors using MDF}

\begin{figure}[ht]
\centerline{\psfig{file=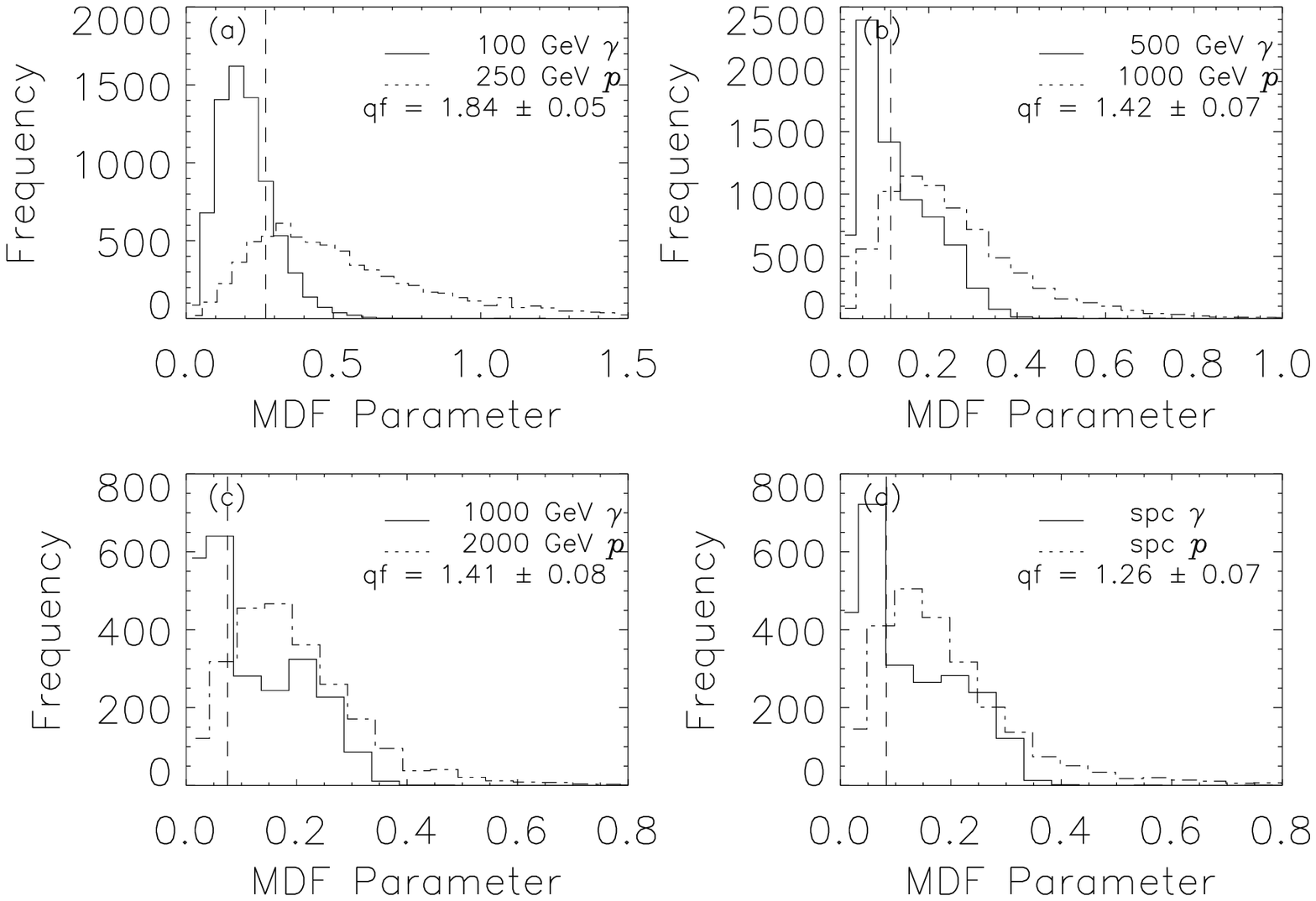,height=12cm,width=12cm }}
\caption{
Distributions of MDF for $\gamma-$ray and proton (dotted line) primaries of
three sets of energies as in figure \ref{rad_ldf}. The dashed vertical lines 
show the threshold values chosen to yield maximum quality factors in each case.
The number of showers simulated in each case are same that mentioned in the
caption of figure \ref{rad_mdf}.} \label{qf_mdf}
\end{figure}

\begin{table}[ht]
\caption{Quality factors from medium range density fluctuations for  primaries
of various energies  incident vertically at the top of the atmosphere.}
\label{tab_mdf}
\vskip 0.25cm
\begin{tabular}{lllc}
\hline
Primary & Threshold  &  Quality Factor  &  Accepted  \\
energy  &  MDF       &                  &  Fraction  \\
\hline
\hline
100 $GeV~\gamma-$rays &  0.27  & 1.84 $\pm$ 0.05  & 0.721 \\
and 250 $GeV$ protons  &  &&0.153\\
\hline
500 $GeV~\gamma-$rays & 0.12  & 1.42 $\pm$ 0.07 & 0.478 \\
and  1 $TeV$ protons &&&0.109\\
\hline
1 $TeV~\gamma-$rays &  0.08 & 1.41 $\pm$ 0.08 & 0.353 \\
and 2 $TeV$ protons &&&0.0625\\
\hline
Spectrum of $\gamma $-rays & 0.083 & 1.26 $\pm$ 0.07 & 0.37  \\
and protons &&& 0.086 \\
\hline
\end{tabular}
\end{table}

Figure \ref{qf_mdf} shows the distributions of MDF for both $\gamma $-ray and
proton primaries of various energies as indicated.  The quality factors are 
listed in table \ref{tab_mdf}, the format of which is same as that of table
\ref{tab_ldf}. It can be seen from table \ref{tab_mdf} that the
quality factors based on MDF as a parameter fall with increasing energy. 
The threshold values too decrease with increasing primary energy as in the 
case of LDF. The fraction of accepted $\gamma $-rays falls more steeply while 
the fraction of protons rejected increases with increasing primary energy 
unlike LDF. The former is due to the progressive increase in MDF for 
$\gamma-$ray primaries around the hump region with increasing energy. We have
included only those showers with core distances less that 150 $m$ while 
estimating these quality factors. In a wavefront sampling experiment like PACT, 
it is possible to estimate the core position from a spherical fit to the
\v Cerenkov light front\cite{cb02}.

It has been seen that the differences in the MDF values of $\gamma $-ray 
and proton primaries is indeed due to intrinsic differences in the interaction
characteristics of the primary species in the atmosphere unlike LDF. This 
conclusion has been arrived at by computing the quality factors after
removing the contribution to MDF from pure statistical (Poissonian)
fluctuations.
 
\subsection{Flatness Parameter}

Flatness parameter is a quantity $\alpha$ defined below, which is 
proportional to the average variance of the total number of \v Cerenkov photons 
incident on each of the 24 telescopes over the 4 sectors.  It is a long range
parameter which represents a measure of smoothness of the lateral distribution
of \v Cerenkov photons generated by $\gamma $-rays and protons. It is defined
as \cite{vis93}: 

\begin{equation}
\alpha = {{1}\over{N}}\sum_{i=1}^{N}{{(\rho _i-\overline{\rho})^2} \over {\overline{\rho}}}
\end{equation}

\noindent
where $\rho _i$ is the total number of \v Cerenkov photons in the $i^{th}$ 
telescope, while $\overline{\rho}$ is the mean $\rho$ averaged over $N$
telescopes in all the 4 sectors, ($N$ = 24 in the present case).

\subsubsection{Lateral Distribution of the flatness parameter}

\begin{figure}[ht]
\centerline{\psfig{file=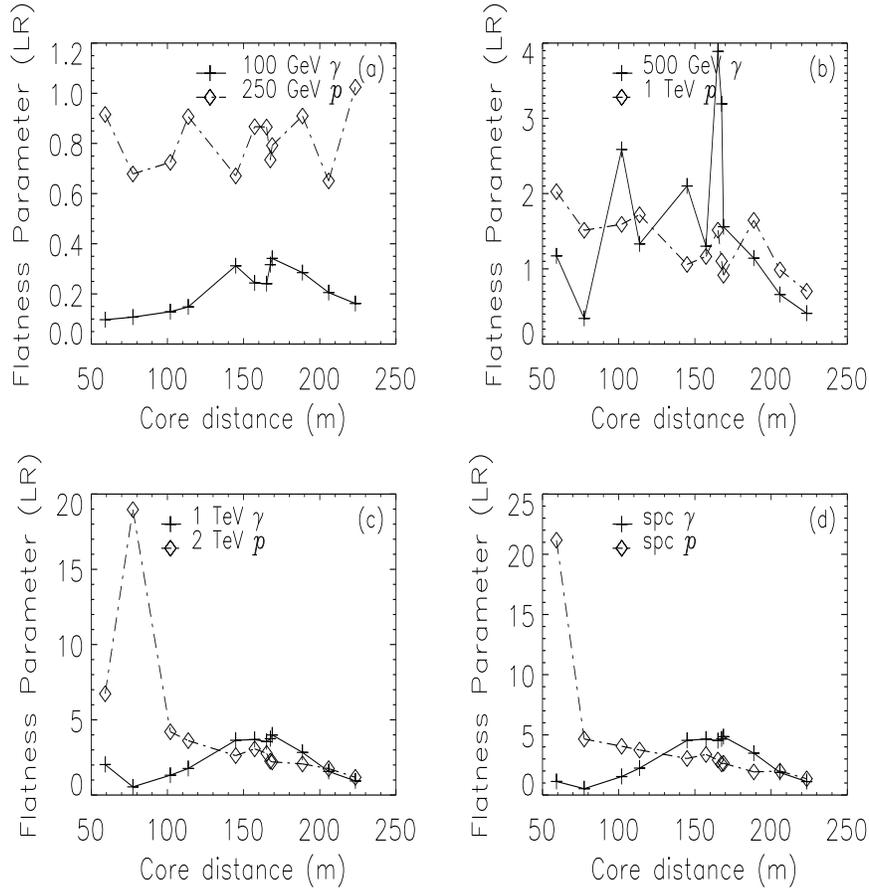,height=12cm,width=12cm }}
\caption{
Radial variations of the mean flatness parameter ($\alpha$) for $\gamma $-ray 
(plus) and proton (diamonds) primaries. $\alpha$ is averaged over 300 events 
each in panels (a) and (b) while it is averaged over 100 events each in panels 
(c) \& (d). The rest of the details are same as in
figure \ref{rad_ldf}. The large fluctuations seen here are statistical. } 
\label{rad_alf}
\end{figure}

Figure \ref{rad_alf} shows the variation of mean $\alpha$ as a function of core 
distance. It can be seen from the figure that $\alpha$ is a core distance 
dependent parameter showing a better separation for $\gamma $-ray and proton
primaries at lower primary energies. In addition, at large core distances and
at higher primary energies its value becomes almost independent of the primary
species. As a result, the sensitivity of $\alpha$ to primary species reduces at
large core distances. This is primarily due to the similarity of the \v Cerenkov
photon lateral distributions from the electromagnetic and hadronic cascades
beyond the hump region. In addition, the number of photons reduces at large
core distances and hence the species independent Poissonian fluctuations
dominate.

\subsubsection{Quality Factors using flatness parameter}

\begin{figure}[ht]
\centerline{\psfig{file=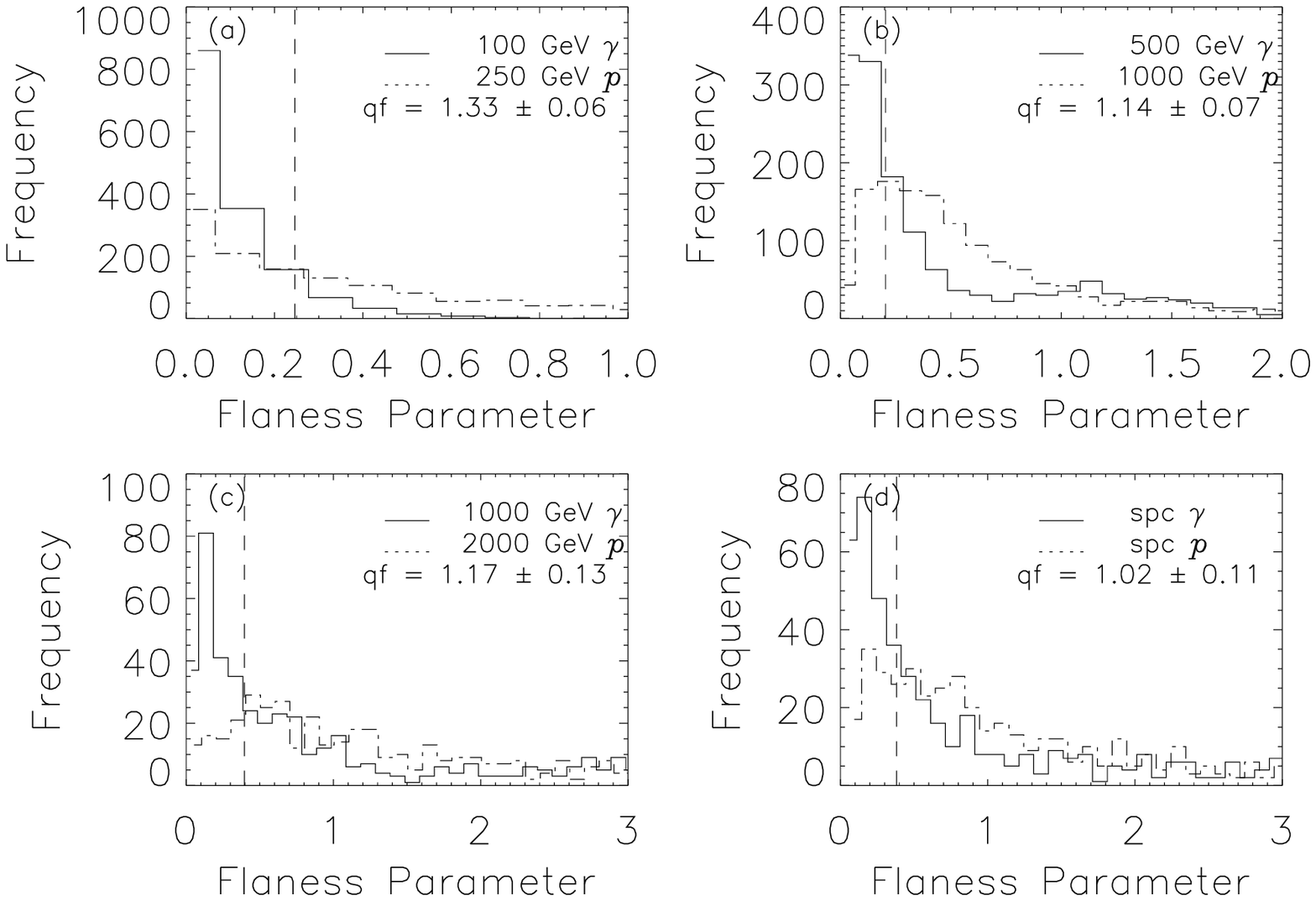,height=12cm,width=12cm }}
\caption{
Distributions of the flatness parameter ($\alpha$) for $\gamma $-ray (plus)
and proton (diamonds) primaries of energies (a) 100 $GeV$ \& 250 $GeV$ (b) 500
$GeV$ \& 1 $TeV$ and (c) 1 $TeV$ \& 2 $TeV$ respectively. Panel (d) shows
similar distributions for $\gamma $-rays \& protons of primary energies randomly
chosen from a power law spectrum of slope -2.65 (see text for details).
The dashed vertical lines show the threshold values chosen to yield maximum
quality factors in each case. The number of showers simulated  is same that in 
figure \ref{rad_alf}.} \label{qf_alf}
\end{figure}

\begin{table}[ht]
\caption{Quality Factors from the flatness parameter ($\alpha$) for primaries 
vertically incident at the top of the atmosphere.} \label{tab_alf}
\vskip 0.25cm
\begin{tabular}{lllc}
\hline
Primary & Threshold  &  Quality &  Accepted Fraction \\
Energy  &  $\alpha$  &  Factor  &  \\
\hline
\hline
100 $GeV~\gamma-$rays &  0.25  & 1.33 $\pm$ 0.06  & 0.836 \\
and 250 $GeV$ protons  &  &&0.395\\
\hline
500 $GeV~\gamma-$rays & 0.20  & 1.14 $\pm$ 0.07 & 0.397 \\
and  1 $TeV$ protons &&&0.121\\
\hline
1 $TeV~\gamma-$rays &   0.4 & 1.17 $\pm$ 0.13 & 0.370 \\
and 2 $TeV$ protons &&&0.100\\
\hline
Spectrum $\gamma $-rays & 0.38 & 1.02 $\pm$ 0.11 & 0.398 \\
&&& 0.152 \\
\hline
\end{tabular}
\end{table}

Figure \ref{qf_alf} shows the distributions of $\alpha$ for $\gamma-$ray and
proton primaries of various energies. The distributions for proton primaries 
are often more skewed relative to that of $\gamma-$ray primaries.
As a result, this parameter can be used to distinguish between the two types of
primaries. The values of the quality factors estimated from these distributions
for primaries of various energies are listed in table \ref{tab_alf}. Also listed
are the fraction of $\gamma-$rays and protons retained after rejecting showers
with $\alpha$ values larger than the threshold values for each set of primary
energies. The fraction of protons that are rejected by using $\alpha$ as a
parameter increases rapidly with increasing energy reaching nearly 90\% at
higher energies. At the same time the fraction of $\gamma-$rays retained
after rejection also decreases to $\sim 40\%$ at higher energies. The 
combined effect of the two energy dependent trends is that the quality factors
decrease with increasing primary energy. A core distance cut of 150 $m$ is used 
while estimating the quality factors.

$\alpha$ can be estimated for a single sector too ($\alpha _s$). It has been 
found that this short range flatness parameter too behaves the same way as the
4-sector or long range flatness parameter except that the quality factors 
estimated using $\alpha _s$ are less by about 10\%.
 
As mentioned before, $\alpha$ is a parameter proportional to the statistical
variance of the total number of \v Cerenkov photons incident on each telescope 
of the array. Higher statistical moments like the skewness and kurtosis too 
have been tested for their sensitivity to primary species. It was found that 
$\alpha$ is the most sensitive among them.

\section{Effect of the telescope field of view on quality factors}

The opening angle of a non-imaging \v Cerenkov telescope is generally limited 
by placing a circular mask
at the focal point in front of the photo-cathode.  This limits the arrival
angle of the photons reaching the photo-cathode. In the absence of a mask the
opening angle is limited by the photo-cathode diameter. In other words, the
limiting
mask is expected to prevent the arrival of photons at large angles. This
effectively results in a reduction in the mean
arrival time as well as an increase in the average production height of 
\v Cerenkov photons\cite{vc01}. 
Table \ref{tab_msk} lists the quality factors (column 4) for each type of 
parameter (column 2) and the corresponding fractions of $\gamma $-rays 
(column 5)
and protons (column 6) retained after applying the cuts as per the threshold
values listed in column 2. The quality factors have been estimated for 4
different focal point mask diameters (column 1). The results are based on 100
showers each simulated for $\gamma $-rays and protons of primary energies 
500 $GeV$ and 1 $TeV$ respectively. It may be noted from the 
table that the quality factors for all the three types of parameters improve 
with
decreasing mask diameter without losing $\gamma $-ray signal.\footnote{This
apparent improvement in quality factor is not due to fixed energies of protons
and $\gamma-$rays. A similar improvement in quality factors was seen when 
the primary energies of protons and $\gamma-$rays were chosen
randomly from a powerlaw spectrum of slope -2.65.} This demonstrates
that the primary species dependent differences in the three parameters arise
mainly due to the intrinsic differences in the cascade development by the
pure electromagnetic and proton primaries rather than the different angular 
distributions of \v Cerenkov photons at the observation altitude.

\begin{table}[ht]
\caption{Quality factors from density fluctuations for showers initiated by 500
$GeV~\gamma-$rays and 1 $TeV$ protons incident vertically at the top of the
atmosphere when focal point masks of various sizes are used.} \label{tab_msk}
\vskip 0.25cm
\begin{tabular}{llllcc}
\hline
Mask & Parameter  & Threshold  &  Quality &  \multicolumn{2}{c}{Accepted}  \\
diameter        &      & value &  factors &  \multicolumn{2}{c}{Fraction} \\
\cline{5-6}
(FWHM)   &      &       &          &  $\gamma-$rays &  Protons \\
\hline
\hline
         & LDF      & 0.18 & 1.53 $\pm$ 0.02  & 0.838 & 0.302 \\
\cline{2-6}
5$^\circ$& MDF      & 0.12 & 1.47 $\pm$ 0.07  & 0.471 & 0.103 \\
\cline{2-6}
         & $\alpha$ & 0.17 & 1.23 $\pm$ 0.09  & 0.334 & 0.074 \\
\hline
         & LDF      & 0.18 & 1.58 $\pm$ 0.02  & 0.843 & 0.285 \\
\cline{2-6}
4$^\circ$& MDF      & 0.09 & 1.50 $\pm$ 0.09  & 0.347 & 0.053 \\
\cline{2-6}
         & $\alpha$ & 0.19 & 1.25 $\pm$ 0.15  & 0.350 & 0.078 \\
\hline
         & LDF      & 0.19 & 1.66 $\pm$ 0.03  & 0.885 & 0.285 \\
\cline{2-6}
3$^\circ$& MDF      & 0.11 & 1.60 $\pm$ 0.09  & 0.437 & 0.075 \\
\cline{2-6}
         & $\alpha$ & 0.21 & 1.37 $\pm$ 0.16  & 0.382 & 0.078 \\
\hline
         & LDF      & 0.20 & 1.77 $\pm$ 0.03  & 0.709 & 0.160 \\
\cline{2-6}
2$^\circ$& MDF      & 0.10 & 2.11 $\pm$ 0.17  & 0.308 & 0.021 \\
\cline{2-6}
         & $\alpha$ & 0.18 & 1.38 $\pm$ 0.19  & 0.302 & 0.048 \\
\hline
\end{tabular}
\end{table}

%\begin{table}[ht]
%\caption{Quality factors from density fluctuations for showers initiated by 
%$\gamma-$rays and  protons (of primary energies randomly chosen from a power law
%spectrum of slope -2.65) incident vertically at the top of the atmosphere when
%focal point masks of various sizes are used.} \label{tab_msk_spc}
%\vskip 0.25cm
%\begin{tabular}{llllcc}
%\hline
%Mask & Parameter  & Threshold  &  Quality &  \multicolumn{2}{c}{Accepted}  \\
%diameter        &      & value &  factors &  \multicolumn{2}{c}{Fraction} \\
%\cline{5-6}
%(FWHM)   &      &       &          &  $\gamma-$rays &  Protons \\
%\hline
%\hline
%         & LDF      & 0.18 & 1.53 $\pm$ 0.02  & 0.838 & 0.302 \\
%\cline{2-6}
%5$^\circ$& MDF      & 0.12 & 1.47 $\pm$ 0.07  & 0.471 & 0.103 \\
%\cline{2-6}
%         & $\alpha$ & 0.27 & 1.23 $\pm$ 0.16  & 0.306 & 0.062 \\
%\hline
%         & LDF      & 0.18 & 1.58 $\pm$ 0.02  & 0.843 & 0.285 \\
%\cline{2-6}
%4$^\circ$& MDF      & 0.09 & 1.50 $\pm$ 0.09  & 0.347 & 0.053 \\
%\cline{2-6}
%         & $\alpha$ & 0.26 & 1.47 $\pm$ 0.21  & 0.302 & 0.042 \\
%\hline
%         & LDF      & 0.19 & 1.66 $\pm$ 0.03  & 0.885 & 0.285 \\
%\cline{2-6}
%3$^\circ$& MDF      & 0.11 & 1.60 $\pm$ 0.09  & 0.437 & 0.075 \\
%\cline{2-6}
%         & $\alpha$ & 0.23 & 2.12 $\pm$ 0.39  & 0.3   & 0.02  \\
%\hline
%         & LDF      & 0.20 & 1.77 $\pm$ 0.03  & 0.709 & 0.160 \\
%\cline{2-6}
%2$^\circ$& MDF      & 0.10 & 2.11 $\pm$ 0.17  & 0.308 & 0.021 \\
%\cline{2-6}
%         & $\alpha$ & 0.33 & 1.51 $\pm$ 0.39  & 0.31  & 0.042 \\
%\hline
%\end{tabular}
%\end{table}

\begin{figure}[ht]
\centerline{\psfig{file=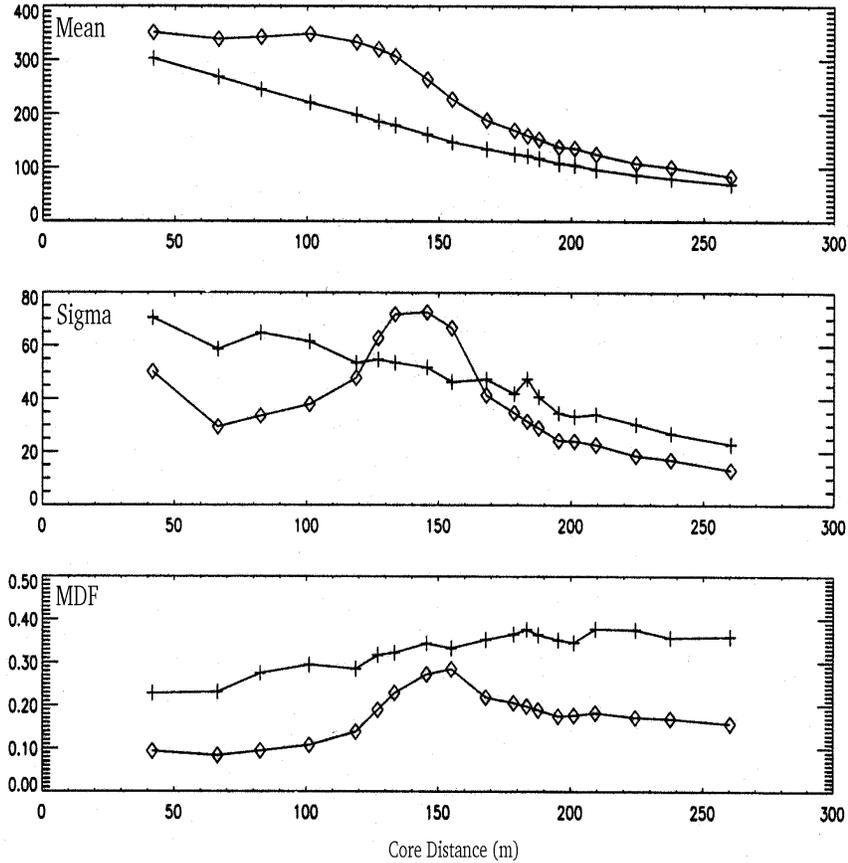,height=12cm,width=12cm }}
\caption{
A plot of the radial variations of the mean photon density (top panel), 
standard deviation (middle panel) and MDF (bottom panel) when no focal point
mask is used. The two curves in each panel correspond to
$\gamma $-rays (diamond) of 500 $GeV$ and protons (plus) of energy
1 $TeV$ incident vertically at the top of the atmosphere. }
\label{rad_mdf_nsk}
\end{figure}

\begin{figure}[ht]
\centerline{\psfig{file=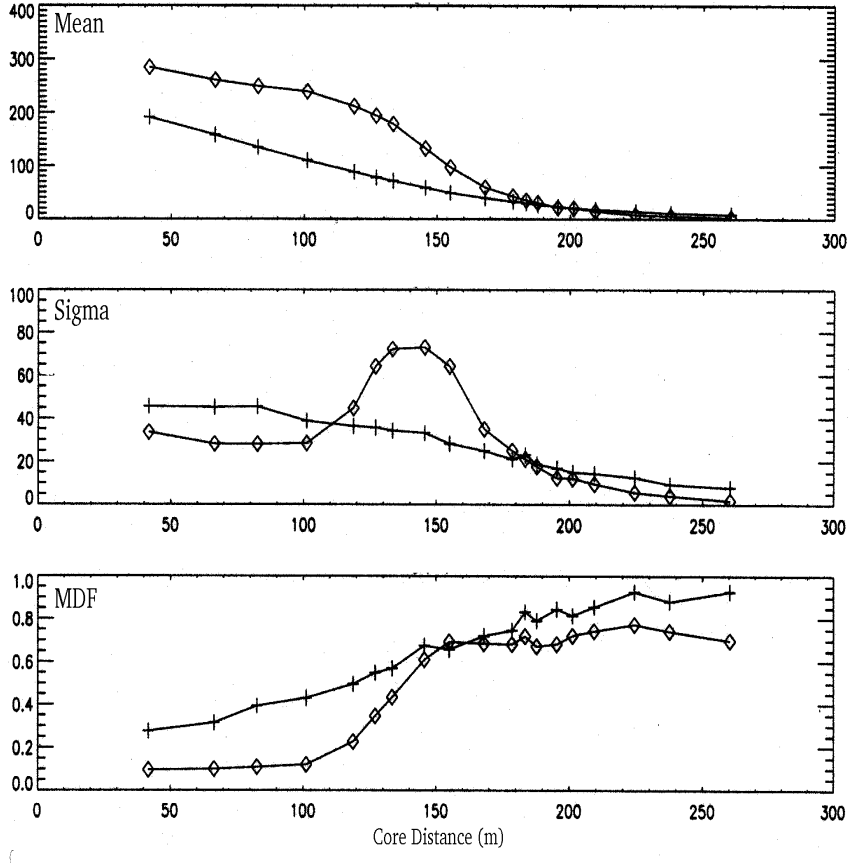,height=12cm,width=12cm}}
\caption{
A plot of the radial variation of the mean value, standard deviation and MDF
when a focal point mask of $4^\circ$ (FWHM) is in place. The 
rest of the details are same as in figure \ref{rad_mdf_nsk}.}
\label{rad_mdf_msk}

\end{figure}

In order to understand the effect of the focal point mask on the density 
parameters we will examine the radial variation of the components of one of
the parameters (MDF) without and with  
a focal point mask in place, as shown in figures \ref{rad_mdf_nsk} and 
\ref{rad_mdf_msk} respectively. From a comparison of the two figures, it can be 
easily seen that the mean photon densities  fall more steeply when focal point 
mask is used while the standard deviation
changes to a lesser extent. This effect is more pronounced for proton primaries
than for 
$\gamma $-ray primaries. As a result, the MDF values increase relatively for
proton primaries thus increasing the separation between the two species which
in turn improve the quality factors when focal point masks are used, as 
observed above. Even though we take MDF
as a typical example to illustrate the effect of focal point mask, the other
parameters too are similarly affected qualitatively.
 
\section{Quality Factors for inclined showers}

At larger zenith angles enhanced attenuation of \v Cerenkov light and increased distance from the
shower maximum raise the energy thresholds of the primary that are detected by
an atmospheric \v Cerenkov telescope. After the shower maximum $\gamma -$ray
induced showers attenuate progressively faster with atmospheric depth than do
the hadronic showers. As a result one would expect a zenith angle dependence
on the sensitivity of the parameters studied here.
Qualitatively speaking inclined showers at a given altitude behave similar to
vertical showers of same primary energy at a lower altitude. Consequently one 
would expect the
quality factor to improve with zenith angle. Quality factors based on some of
the timing parameters show an improvement at inclined direction \cite{vc01}.
The species sensitive imaging parameters like the azwidth on the other hand, 
have been shown to be much less sensitive for primaries incident at angles 
$\ge 30^\circ$ \cite{we89}.

Table \ref{tab_incl} summarizes the quality factors estimated for $\gamma -$ray
and proton primaries of energy 500 $GeV$ and 1 $TeV$ respectively, incident at 
$30^\circ$ to the vertical at the top of the atmosphere. These are based on 100
showers simulated for each type of primary. These quality factors when
compared with those for vertical showers as listed in tables \ref{tab_ldf}, 
\ref{tab_mdf} and 
\ref{tab_alf} show a marked improvement for $\alpha$ and MDF as parameters 
while it doesn't change significantly for LDF as a parameter.

\begin{table}[ht]
\caption{Quality Factors estimated for 500 $GeV~\gamma $-rays and 1 $TeV$
protons incident at $30^\circ$ to the vertical at the top of the atmosphere.
The quality factors for three different parameters estimated with no focal 
point mask are shown.} \label{tab_incl}
\vskip 0.25cm
\begin{tabular}{lllcc}
\hline
Type of & Threshold  &  Quality & \multicolumn{2}{c}{Accepted Fraction} \\
\cline{4-5}
parameter  &   &  Factor  & $\gamma-$rays  & Protons \\
\hline
\hline
$\alpha $ &  0.35  & 3.28 $\pm$ 0.79  & 0.328 &0.010\\
\hline
MDF & 0.12  & 1.83 $\pm$ 0.08 & 0.617&0.113\\
\hline
LDF &   0.22 & 1.39 $\pm$ 0.02 & 0.850 &0.377\\
\hline
\end{tabular}
\end{table}
\vskip 0.25cm
\section{Quality Factors for heavy primaries} 

Each of the three species sensitive parameters under study here are applied to
$He$ and $Fe$ primaries as well. Tables \ref{tab_he} and \ref{tab_fe} 
summarize the results 
for 100 simulated showers each of 1 $TeV~\gamma $-rays 2.5 $TeV~He$ and 10 
$TeV~Fe$ nuclei
incident vertically at the top of the atmosphere. The quality factors may be 
compared with those for protons as listed in tables \ref{tab_ldf},
\ref{tab_mdf} and \ref{tab_alf} which show that both $He$ and $Fe$ nuclei may 
be more
easily discriminated against $\gamma $-rays despite losing a higher fraction 
of $\gamma $-rays in the process. The fraction of cosmic rays retained after
applying the cut is smaller in the case of heavier primaries for all the 
three density based parameters studied here.

\begin{table}[ht]
\caption{Quality Factors estimated for 1 $TeV~\gamma $-rays and 2.5 $TeV$
{\it He} nuclei incident vertically at the top of the atmosphere.
The quality factors for three different parameters estimated with no focal
point mask are shown.} \label{tab_he}
\vskip 0.25cm
\begin{tabular}{lllcc}
\hline
Type of & Threshold  &  Quality & \multicolumn{2}{c}{Accepted Fraction} \\
\cline{4-5}
parameter  &   &  Factor  & $\gamma-$rays  & $He$ nuclei \\
\hline
\hline
$\alpha $ &  0.57  & 1.11 $\pm$ 0.06  & 0.303 &0.074\\
\hline
MDF & 0.07  & 1.68 $\pm$ 0.12 & 0.312&0.035\\
\hline
LDF &   0.11 & 1.46 $\pm$ 0.02 & 0.738 &0.256\\
\hline
\end{tabular}
\end{table}

\begin{table}[ht]
\caption{Quality Factors estimated for 1 $TeV~\gamma $-rays and 10 $TeV$
{\it Fe} nuclei incident vertically at the top of the atmosphere.
The quality factors for three different parameters estimated with no focal
point mask are shown.} \label{tab_fe}
\vskip 0.25cm
\begin{tabular}{lllcc}
\hline
Type of & Threshold  &  Quality & \multicolumn{2}{c}{Accepted Fraction} \\
\cline{4-5}
parameter  &   &  Factor  & $\gamma-$rays  & $Fe$ nuclei \\
\hline
\hline
$\alpha $ &  0.57  & 1.89 $\pm$ 0.15  & 0.301 &0.025\\
\hline
MDF & 0.07  & 2.23 $\pm$ 0.18 & 0.331&0.022\\
\hline
LDF &   0.10 & 1.34 $\pm$ 0.02 & 0.659 &0.243\\
\hline
\end{tabular}
\end{table}

\section{Discussions}

The subject of intrinsic inter-shower fluctuations has been dealt at length by 
Chitnis and Bhat \cite{vc98}. Here what we are addressing is the techniques of 
exploiting intra-shower fluctuations in \v Cerenkov photon density at the
observation level. It is well known that in the case of hadronic primaries
large fluctuations in the number of secondary particles created during the
hadron multi-particle production is the main reason for larger fluctuations
relative to that in photon primaries.

As can be seen from the tables \ref{tab_ldf},\ref{tab_mdf} and \ref{tab_alf} 
that the quality factors fall with increasing primary energy. This is
consistent with our previous
study \cite{vc98} where it was observed that the intra-shower density 
fluctuations decrease monotonically with increasing primary energy, thus 
reducing the distinguishability  of $\gamma $-rays from protons. It may be 
interesting to note here that the fraction of protons rejected using these
parameters increase with increasing primary energy at the cost of losing more
$\gamma $-ray signal.

%\begin{table}[ht]
%\caption{Quality factors when $alpha$,MDF \& LDF applied in tandem to the same 
%dataset for  primaries of various energies
%incident vertically at the top of the atmosphere.} \label{tab_alf_mdf_ldf}
%\vskip 0.25cm
%\begin{tabular}{lcccc}
%\hline
%Mask &  \multicolumn{4}{c}{Quality Factors}    \\
%\cline{2-5}
%Diameter& 100 $GeV~\gamma$ & 500 $GeV~\gamma$ & 1 $TeV~\gamma$ & Spectrum of $\gamma$\\
%(deg. FWHM) & 250 $GeV$ $p$ & 1 $TeV$ $p$   & 2 $TeV$ $p$ &  and $p$\\  
%\hline 
%\hline
%No mask & 2.22 $\pm$ 0.06 & 1.53 $\pm$ 0.08 &1.54 $\pm$ 0.10 &1.29 $\pm$ 0.07 \\
%\hline
%5.0 & 2.28 $\pm$ 0.07 & 1.57 $\pm$ 0.09 &1.62 $\pm$ 0.10 &1.29 $\pm$ 0.07 \\
%\hline
%4.0 &  2.34 $\pm$ 0.07 & 1.59 $\pm$ 0.11 &1.62 $\pm$ 0.11 &1.29 $\pm$ 0.07 \\
%\hline
%3.0 & 2.37 $\pm$ 0.07 & 1.84 $\pm$ 0.12 &1.76 $\pm$ 0.13 &1.29 $\pm$ 0.07 \\
%\hline
%2.0 & 2.52 $\pm$ 0.09 & 2.73 $\pm$ 0.29 &2.03 $\pm$ 0.17 &1.29 $\pm$ 0.07 \\
%\hline
%\end{tabular}
%\end{table}

\begin{figure}[ht]
\centerline{\psfig{file=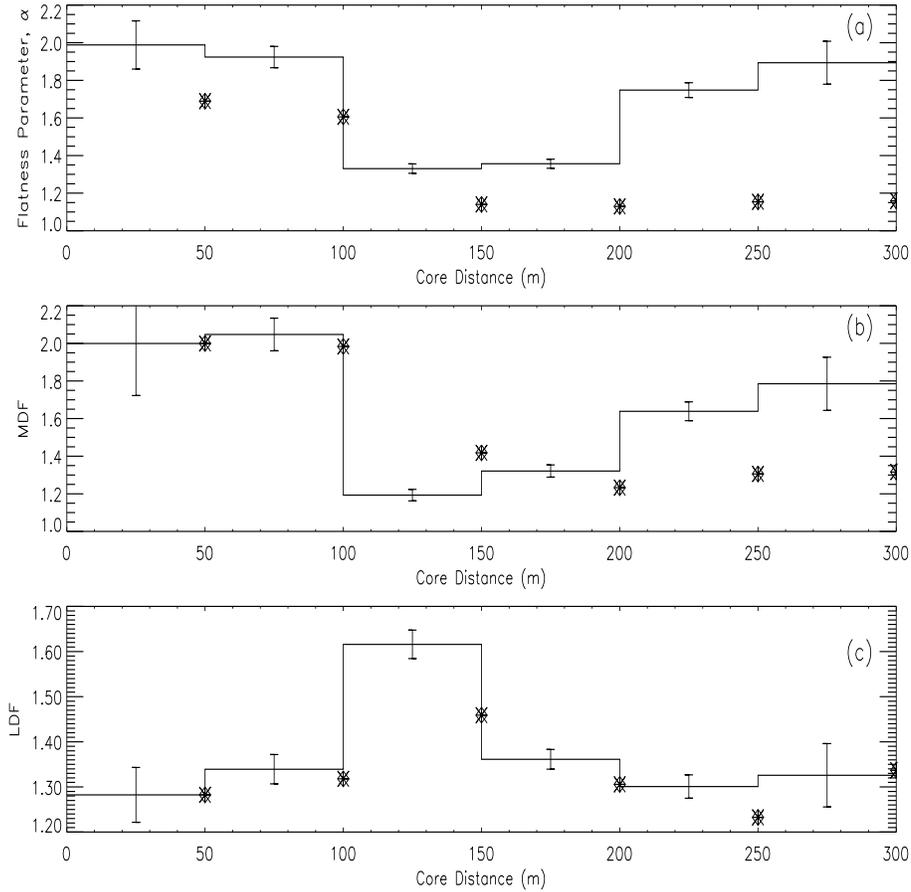,height=12cm,width=12cm }}
\caption{
A plot showing the dependence of the three quality factors as a function of
core distance cuts, both integral and differential. The histogram shows
the differential core distance dependence of quality factors in units of 50 $m$
for each of the parameters (a) $\alpha$, (b) MDF and (c) LDF. The points
(asterisks) show the quality factors when the core distances are
chosen in the integral mode in units of 50 $m$. The quality factors are
estimated for 500 $GeV$ $\gamma $-rays and 1 $TeV$ protons where 100 showers 
were simulated for each primary species.}
\label{rad_qf}

\end{figure}

It has been seen in figures \ref{rad_ldf},\ref{rad_mdf} and \ref{rad_alf} that 
the \v Cerenkov photon density fluctuations are strong
functions of the core distance as the photons received at various core
distances are produced at different altitudes, {\it i.e.} at different stages in
the cascade development \cite{hil96,vc98}. As a result, we see a dependence of
the quality factors on the core distance. We have used a uniform cut of 
150 $m$ on the core distance in all the present studies (except for inclined
showers since the \v Cerenkov light pool generated by them is azimuthally
asymmetric) since no showers of primary energy $\le 2~TeV$ with
impact parameter above 150 $m$ will generate a PACT trigger. 
However it is important to know the effect of this cut on the quality factors
studied here. Figure \ref{rad_qf} shows the variation of quality factors based
on (a) $\alpha$, (b) MDF and (c) LDF as a function of core distance cuts
(increasing in units of 50 $m$, 0-50 $m$, 0-100 $m$ {\it etc.}, shown as
points) used for primary energies of
500 $GeV$ and 1 $TeV$ for $\gamma $-rays and protons respectively. The quality
factors from $\alpha$ and MDF fall continuously when longer core distance
events are included while that from LDF shows a maximum when the core distance
cut is 150 $m$.

Also shown in the figure are the quality factors when only showers with
differential impact parameter ranges 0-50 $m$, 50-100 $m$ {\it etc.,} are
included (histogram). The quality factors from MDF and $\alpha$ show a minimum
for showers with impact parameters around the hump region ({\it 100-150 $m$}
for Pachmarhi altitude). This is primarily due to the fact that the absolute
values of MDF and $\alpha$ exhibit minimum separation around the hump region
as seen in figures \ref{rad_mdf} and \ref{rad_alf} respectively. The quality
factors based on LDF show a maximum in the same core distance range. This is a
direct consequence of the hump in the case of $\gamma $-ray primaries. 

In practice one can estimate the primary energy from which we can estimate the
maximum possible core distance. This would enable us to make an optimum
estimate of the nature of the primary. Alternately one can estimate the core
position of each shower by using the curvature of the shower front\cite{cb02},
as mentined before, which would help in deciding the optimum cut on the
core distance for the data set.

If the three types of photon density based parameters investigated in the
present study are independent,\footnote{LDF and MDF are expected to be
independent since they use exclusive information content while $\alpha$ and MDF
could be related. It has been found that the improvement resulting from the
tandem application of $\alpha$ over and above LDF \&  MDF is marginal as
expected.} then one can use them in tandem to improve the hadron
rejection efficiency even further, a procedure similar to the multi-dimensional
shower image analysis \cite{ack89}. After applying the
cuts based on the three parameters, to the same data-set in tandem the resulting
quality factors are 2.22 $\pm$ 0.06, 1.53 $\pm$ 0.08, 1.54 $\pm$ 0.10 for the 
three distinct primary energies studied here {\it viz.} 100 \& 250 $GeV$, 0.5
\& 1.0 $TeV$ and 1.0 \& 2.0 $TeV$ respectively. Thus using the \v Cerenkov 
photon density based parameters alone one is able to reject more than 90\% of 
background protons.

The parameter $\alpha$ has been used by the CELESTE group to reject nearly
80\% of the proton primaries from their data from the Crab Nebula. From their
simulation studies a quality factor of 1.6 has been obtained while the 
$\gamma $-ray energy threshold of CELESTE is $\sim 50~GeV$. From the present
studies we have estimated a quality factor of 1.33 for primary energies of 
100 $GeV$ and 250 $GeV$ for $\gamma $-rays and protons respectively for
Pachmarhi altitude (table \ref{qf_alf}. Considering their lower energy
threshold (50 $GeV$), their quality factor is consistent with ours \cite{nau00}.

\section{Conclusions}

In this work we have examined the feasibility of improving the signal to
noise ratio using the differences in the fluctuations of \v Cerenkov photon
distribution in the light pool generated by $\gamma $-ray and proton initiated
showers. The estimates of quality factors are relevant to the configuration
of PACT which uses the wavefront sampling technique. However the quality factors
using the parameters studied here can be easily optimized
to any array configuration of \v Cerenkov telescopes.

Various shower characteristics like the image shape, distribution of light on
the ground, time profile \& structure, spectrum, polarization and the UV 
content in the \v Cerenkov light have been suggested in the literature for
hadron discrimination.  It has been verified experimentally 
that the shape and orientation of the \v Cerenkov images
can reject more than 99\% of the background \cite{hil96}.
Similarly, from our earlier studies on
\v Cerenkov photon temporal properties \cite{vc01} and the present 
investigations on their photometric properties we can conclude that one can 
efficiently reject more than 90\% of the cosmic ray background while adopting 
wavefront sampling technique.
 
\acknowledgements{We would like to acknowledge the fruitful discussions with
and helpful suggestions from Profs. K. Sivaprasad, P. R. Vishwanath and B. S.
Acharya during the present work.}

\end{article}
\end{document}